\documentclass{webofc}
\pdfoutput=1
\usepackage[varg]{txfonts}   
\usepackage{color}
\usepackage{lineno}
\begin{document}
\title{Summary of the plenary sessions at European Space Weather Week 15: space weather users and service providers working together now and in the future}
\thispagestyle{plain}
\pagestyle{plain}
%
%

\author{\firstname{Suzy} \lastname{Bingham}\inst{1}\fnsep\thanks{\email{suzy.bingham@metoffice.gov.uk}} \and
        \firstname{Sophie A.} \lastname{Murray}\inst{2,3} \and
        \firstname{Antonio} \lastname{Guerrero}\inst{4} \and
        \firstname{Alexi} \lastname{Glover}\inst{5} \and
        \firstname{Peter} \lastname{Thorn}\inst{1}
}

\institute{Met Office, Exeter, United Kingdom
\and
           School of Physics, Trinity College Dublin, Dublin, Ireland
\and
           School of Cosmic Physics, Dublin Institute for Advanced Studies, Dublin, Ireland
\and
           Universidad de Alcal\'{a}, Alcal\'{a} de Henares, Madrid, Spain
\and 
           European Space Agency, Darmstadt, Germany       
          }

\abstract{
  During European Space Weather Week 15 two plenary sessions were held to review the status of operational space weather forecasting. The first session addressed the topic of working with space weather service providers now and in the future, the user perspective. The second session provided the service perspective, addressing experiences in forecasting development and operations. Presentations in both sessions provided an overview of international efforts on these topics, and panel discussion topics arising in the first session were used as a basis for panel discussion in the second session. Discussion topics included experiences during the September 2017 space weather events, cross domain impacts, timeliness of notifications, and provision of effective user education. Users highlighted that a \textit{severe} space weather event did not necessarily lead to severe impacts for each individual user across the different sectors. Service providers were generally confident that timely and reliable information could be provided during severe and extreme events, although stressed that more research and funding were required in this relatively new field of operational space weather forecasting, to ensure continuation of capabilities and further development of services, in particular improved forecasting targeting user needs. Here a summary of the sessions is provided followed by a commentary on the current state-of-the-art and potential next steps towards improvement of services.
}
\maketitle
\section{Introduction}
\label{meet_intro}
During the week of the 5th November 2018, the fifteenth European Space Weather Week (ESWW15) was held in Leuven, Belgium. There were two plenary sessions, one oriented to users and another one oriented to operations and services. The first plenary session was titled `Working with space weather services now and in the future'\footnote{http://www.stce.be/esww15/program/session\_details.php?nr=5}, and later in the week a second session took place titled `Space weather operations and services'\footnote{http://www.stce.be/esww15/program/session\_details.php?nr=10}. Following a mix of invited and contributed talks, panel sessions were held in the final part of each oral session focusing on experiences during the space weather events of September 2017\footnote{http://www.stce.be/esww15/program/Sept\_2017\_activity\_summary.pdf} (see Redmon et al. 2018 for an overview of the 2017 events). In preparation for the plenary sessions, a number of talking points were provided to presenters and panellists to encourage discussion within the sessions, which are outlined in the Appendix. From these points several recurring topics emerged throughout the sessions that identified important areas of improvement within the space weather community, not just related to forecast output but also regarding communication between researchers, operators, and end-users. These topics are discussed further in the Commentary.

Poster sessions were held alongside each session covering numerous subjects from best practices in transitioning models from research to operations, to ESA's Virtual Space Weather Modelling Centre, part of the ESA SSA Space Weather Network\footnote{http://swe.ssa.esa.int}, to forecast verification.

The September 2017 events took place over a ten day period of enhanced solar activity between 4th – 13th September 2017 originating from active region (AR) 12673, as numbered by the NOAA Space Weather Prediction Center (SWPC). Activity began with the AR rapidly developing from a simple to a complex configuration between the 3rd and 5th September. The AR produced four X-class solar flares, including an X9.3 eruption on the 6th September that was the largest flare recorded in solar cycle 24 and the brightest recorded since an X17 flare in September 2005.  Three Earth-directed coronal mass ejections (CMEs) erupted leading to a maximum NOAA SWPC geomagnetic storm scale level of G4, \textit{severe} (Kp 8.3). An X8.2 flare on the 10th September resulted in a solar energetic particle (SEP) event and a ground level enhancement (GLE). 

The September 2017 activity was informally referred to as 'severe', although the severe threshold was crossed only for the NOAA SWPC geomagnetic storm scale (G). During the events, a maximum level of \textit{strong} was reached for the radio blackout (R) and solar energetic particle event (S) scales. There was therefore increased risk for systems susceptible to strong or severe conditions, including: wide-area blackout of high-frequency (HF) radio communication; degradation of low frequency navigation signals; single-event upsets (SEUs); surface charging and tracking problems in spacecraft operations; widespread voltage control problems in power systems. 

It is worth noting that \textit{extreme} is the highest on the scale, and such events are generally estimated to occur every 100 years (see Riley et al. 2018a for a summary of current understanding of extreme event characteristics and impacts). An extreme event could include the following characteristics: ultra intense X-ray flare; fast (> $\sim$3,000 km/s) and massive ($\sim$5x10$^{13}$ kg) CME with large southward Bz (< $\sim$-100 nT); large fluxes of particles; large variation in geomagnetic parameters such as Dst (< $\sim$-600 nT); sudden ionospheric disturbances; strong geomagnetically-induced currents (GICs). Potential impacts due to an extreme event include satellite failures and power outages, as well as cascading problems from these primary impacts such as food and water shortage (Oughten et al. 2017).

During the sessions some speakers used 'validation' and some used 'verification' to mean the evaluation of a model or forecast, for example to measure performance such as accuracy. In this review we use 'verification' throughout for consistency.

\section{Users Session}
\subsection{Summary of presentations}
In the course of the first session, the topic of which was the space weather user perspective, presentations particularly focused on experiences and concerns from the satellite sectors (SES\footnote{https://www.ses.com/}, EUTELSAT\footnote{https://www.eutelsat.com/en/home.html} and ESSP-SAS\footnote{https://www.essp-sas.eu/}). To be operationally prepared, geostationary orbit (GEO) satellite operators indicated that they had a particular interest in space weather radiation effects such as single event effects (SEEs), total dose increases, surface charging, internal charging, and electrostatic discharge. Operators emphasised that information on the current and future radiation environment was important but that access to historical data was also required to allow operators to understand the survivability of a spacecraft. Technical standards and guidelines are available to spacecraft engineers in order to avoid and minimise hazardous effects arising from the space environment, for example, the European Cooperation for Space Standardisation (ECSS) has published a spacecraft charging standard (ECSS 2019) and an assessment of space worst case charging handbook (ECSS 2019a). Similarly, NASA provides guidelines, for example, for space charging effects (NASA 2011) which include typical spacecraft design solutions.  

During the session, the September 2017 events were reported as insignificant to GEO satellites compared to historical activity such as the Halloween Storms of October 2003 and events in March 2012. Space weather can affect satellites in several different ways (see Horne et al. 2013 for a summary). Between the 23rd October and the 6th November 2003, 47 satellites reported malfunctions and the Midori 2 was lost completely (Horne et al. 2013). These anomalies occurred when there were very large changes in the radiation belts and where a solar energetic particle (SEP) event took place at the same time. A \textit{severe} solar radiation storm (S4) was recorded. In March 2012, the Sky Terra 1 satellite reported problems resulting in loss of service for a few days (Horne et al. 2013). This was during a moderate geomagnetic storm but GEO satellite observations indicated multiple substorms and plasma injections which are known to cause surface charging. The highest category solar radiation storm measured was S3, \textit{strong}. Spaceway 3 and GOES 15 also reported outages for a few hours and a few days respectively during March 2012. During the September 2017 events, 100 MeV protons were observed at GEO (see Redmon et al. 2018 table 2 for data source locations) and the SWPC electron event alert threshold, issued when relativistic (>2 MeV) electron flux exceeds 1,000 pfu, was exceeded semi-continuously. The \textit{strong} (S3) solar radiation storm threshold was reached. Jiggens et al. 2019 calculate that the solar particle events (SPEs), characterised by large enhancements of particle radiation fluxes from the Sun, of September 2017 were not particularly concerning in terms of dose effects in spacecraft electronics. Jiggens et al. 2019 do however note several spacecraft anomalies which were observed, for example, a technology demonstration payload on the AlphaSat platform in GEO recorded a higher rate of single bit upsets (SBUs) during the SPE on the 11th September. 

Operators remarked that total dose effects can lead to long term degradation of materials and that the last time significant solar array degradation was detected at GEO due to a single period of events was 15 years ago during the 2003 Halloween Storms. A concern was raised that there had been a long period of relatively benign conditions since 2003 and, as a result, many working in the domain currently may not have hands-on experience of highly disturbed conditions, placing the emphasis on good training being available to support preparedness. 

SEUs in the main computer of a spacecraft have been associated with space weather but operators can correct such effects by a system reset with reportedly no impact on the integrity of the mission. SEUs were reported to have increased significantly during September 2017 but remained within design margins and so no impact or degradation was reported. Operators did report Earth sensor glitches during the September 2017 events but remarked that spacecraft are designed to withstand these. During the session, satellite operators reported that no unmitigated issues arose during the September 2017 events. Operators expressed that satellites were demonstrated to be already robust to space weather events such as those in September 2017 but that service providers should be able to clearly inform them of exceptional events such as a Carrington-type event.  

The final talk of the users session was given by a representative from the UK Government Department for Business, Energy and Industrial Strategy (BEIS\footnote{https://www.gov.uk/government/organisations/department-for-business-energy-and-industrial-strategy}). BEIS reported that there had been no known impacts or damage in the UK due to the September 2017 events. However, it was emphasised that space weather was a significant risk compared to other natural hazards, as several areas of infrastructure could be impacted or lost during an event, such as communications and electricity. 

BEIS said it would ensure that the UK was resilient and able to respond to a space weather event on the basis of proportionate risk mitigation through efforts in the following hazard preparedness phases: Predict, Understand, Prepare, Respond, and Recover. The sectors potentially at risk include power, transport, satellite navigation and timing, telecommunications, central government, and local government. Increasing resilience in the UK has been addressed, for example, through establishing the Met Office Space Weather Operations Centre (MOSWOC) which began 24/7 space weather operational forecasting in 2014; developing protocols between MOSWOC and industries to warn of forthcoming events and to maximise the time for mitigating actions to be taken; increasing understanding of the risk through economic impact studies; ensuring that generic capability planning (including at local level) reflects the impacts of severe space weather; developing key sectors’ resilience to space weather and developing new capability where needed whilst avoiding increasing vulnerability in the future; having in place plans for response and exercising these plans. 

After the risk of severe space weather was added to the UK National Risk Assessment in 2011, a Space Weather Preparedness Strategy was published by the UK Government Department for Business Innovation and Skills in 2015 (BIS 2015). Many of the impacts of space weather are common to other hazards, for example, power loss, interruption to satellite services, transport disruption, and loss of communication, therefore some existing response plans have been used to deal with space weather. One challenge which BEIS specifically highlighted was engagement within the transport sector, noting that avionics could be affected by space weather and that electronics could also be at risk on the ground in the future roll-out of autonomous vehicles. The UK’s approach was to assess each sector separately and, for example, to build upon work within the more established area of aviation to increase awareness of and resilience to the risk of severe space weather across other less aware areas of transport, including rail, road, and maritime. 

\subsection{Panel discussion}
In the panel part of the users session, panellists (L. Libor, S. Magdaleno, D. Pitchford, D. Zamora, and a BEIS representative) reported on experiences during the September 2017 events when space weather data and services were utilised to monitor activity. No damage was reported by the users present. However, degradation in HF communications over the Atlantic, polar regions, and high latitudes were reported at the time and high latitude solar dose rates were reported to be at alert levels (20 $\mu$Sv/hour  at 70,000 feet) on the Enhanced Solar Radiation Alert System (ESRAS; Copeland 2016). The panel discussions highlighted the difference in levels of space weather awareness across the different sectors from spacecraft operators to rail and aviation. Some satellite operators suggested that the September 2017 events should not be classed as \textit{severe}, comparing to those greater impacts which were experienced during the 2003 storms. 

The panel discussed how the space weather community could develop understanding for emerging user groups, one example highlighted was to support the rail community in gathering evidence of space weather impacts on the rail sector. Education and training for users was deemed a necessary requirement although providing these effectively would require careful consideration due to the wide range of users concerned. It was remarked that the priority should be the provision of a course giving an introduction to space weather for the general user, prior to bespoke courses for different sectors. The potential impact of space weather may increase in the future as technology develops in areas such as the roll-out of 5G, large satellite constellations being launched, and the increased use of autonomous vehicles. The panel suggested that space weather should be considered in the design of these systems, noting that education and training in space weather may be needed to support this. A part of a risk assessment is the assessment of vulnerability; service providers noted that information about potential vulnerabilities of affected systems such as satellites is often not available due to this being potentially commercially sensitive information. With more detailed information about potential vulnerabilities, more tailored information could be provided. Where this isn't provided, service providers usually focus on providing information on ongoing or forecast conditions along with indicative information about the type of effects which may be observed. For railways, it was said that it may be possible to share rail technology risk thresholds with service providers in future. 

There was also discussion around the need for timely notifications of severe conditions, with longer lead-times to provide better warnings, in an easy-to-digest format, and with further access to more detailed information when available. Satellite operators remarked that their ideal requirement would be 2-3 weeks of notice of an extreme event. Cadence and an up-to-date and reliable service were important, along with information on the potential development of an event. Operators also emphasised that forecasting of all-clear periods was beneficial, for example, a 3 week window of ‘good weather’ would be a useful piece of information. A database of anomalies would also be useful to correlate space weather activity with impacts. Users described how they must balance mitigating action with the expected impact to their service and the underpinning infrastructure, thus emphasising the need for reliable space weather forecasts. In general, users reported good results so far in building close relationships between forecast centres and individual users, although this was domain dependent and related to the level of awareness of the user.

\section{Services Session}
\subsection{Summary of presentations}
The second session focused on experiences from the space weather operations and service providers perspective, with forecast centre overviews from the Solar Influences Data Analysis Centre (SIDC, Belgium)\footnote{http://sidc.oma.be/}, Norwegian Centre for Space Weather (NOSWE)\footnote{http://www.spaceweather.no}, and Korean Space Weather Centre (KSWC)\footnote{https://spaceweather.rra.go.kr/?lang=en}. The research to operations perspective was explored through talks from the European Commission Ionosphere Prediction Service project (Guyader et al. 2018), Community Coordinated Modeling Center (CCMC) CME Arrival Time and Impact Working Team\footnote{https://ccmc.gsfc.nasa.gov/assessment/topics/helio-cme-arrival.php}, and Application Usability Level (AUL) framework\footnote{https://aerospace.org/sites/default/files/2018-12/AUL\%20Brief.PDF} (Halford et al. 2019). The final presentation was from NOAA SWPC\footnote{https://www.swpc.noaa.gov/} on experiences during the September 2017 events.

SIDC space weather forecasters are active researchers and SIDC plans to transition into becoming a 24/7 operation centre soon. SIDC provides a daily bulletin and alerts and forecasts of flares, geomagnetic storms, and F10.7. It has in-house solar optical and radio ground-based instruments and in-house satellite instrument operations (such as Proba 2 EUV images). It also has access to data (GNSS, geomagnetic, neutron monitoring) from sister institutes and imports routine automated data from external sources. SIDC uses Solar Demon Flare Detection and SoFAST to automatically detect and locate flares (Kraaikamp \& Verbeke 2015), and uses Computer Aided CME Tracking (CACTUS; Robbrecht et al. 2009) to detect and characterise CMEs. An international network of solar radio spectrometers, e-Callisto\footnote{http://www.e-callisto.org}, is used to detect radio bursts. SIDC uses the European heliospheric forecasting information asset (EUHFORIA; Pomoell \& Poedts 2018) for solar wind and CME forecasting. SIDC works with users to understand the type of information they require, what SIDC can offer, how timely information can be delivered and how to communicate the information. SIDC conveyed that there were benefits of having researchers as forecasters, for example, that they have sound scientific knowledge to assess uncommon situations and have knowledge of the latest tools and research insights. On the other hand, the operational tools and practices used could lack consistency with researchers in the role. SIDC explained that the main challenge was maintaining funding streams to support the operational environment and to support projects. Action areas included the streamlining of operational data processing deployments, maintaining and increasing consistency in human operations, and improving understanding in customer needs. A number of ongoing actions were highlighted including developing operator tools and procedures for increasing consistency, as well as increasing customer contacts (which was noted as a slow, iterative process).

NOSWE is a semi-independent unit, part of the Tromso Geophysical Observatory (TGO\footnote{http://www.tgo.uit.no}) and was established in collaboration with the Norwegian Space Centre (NSC\footnote{https://www.romsenter.no/eng/}). NOSWE, like SIDC, has researchers who also take on the role of duty forecaster. About 60\% of NOSWE forecast users are located in Norway and about 10\% in the UK. NOSWE remarked that in terms of space weather, geomagnetic activity had the largest economic impact in the region and therefore NOSWE's primary services addressed the needs of the oil drilling industry and auroral tourism. NOSWE plans to support power grid operators by supplying forecasts and by offering local expertise, for instance NOSWE will be able to discuss with control centre operators the implications of the forecast on power lines and transformers. The Norwegian electricity industry has requested a GIC forecast that is extremely precise and highly reliable with a minimum of 4-5 days of notice before storm arrival at Earth.

Established in 2011, KSWC’s main priorities are research into technology to aid space weather forecasting; observations of solar, geomagnetic and ionospheric activities; and analysis of observation data. KSWC provides forecasts to around 4,500 users in the satellite, communication, defence, aviation, and power sectors. Forecasters use an interplanetary scintillation driven version of the Enlil model (Odstrcil 2003) for solar wind and CME forecasting. KSWC’s Automatic Solar Synoptic Analyzer (ASSA) system provides real-time monitoring and identification of solar phenomena such as sunspot groups and filaments. KSWC aims to improve its flare prediction accuracy through the development of the Post ASSA system which will use cutting edge deep learning techniques instead of the current statistical method based on Bloomfield et al. 2012. KSWC is also developing a model for the prediction of ionospheric variation around Korea by combining data assimilation with a physics-based model to predict ionospheric electron density, along with a domestic ionosphere index. KSWC also described development of two verification systems, one for calculating reliability of 3-day forecasts, and the other for verification of model predictions.

The presentation on the Galileo Ionosphere Prediction Service (IPS) stated that the objective of IPS was to provide monitoring, prediction, and warnings to mitigate the impacts of disruptive events on the operations of several GNSS based application domains. IPS provides solar activity products (monitoring and prediction of flares, CMEs and SEPs), ionospheric activity products (nowcasting and forecasting of total electron content and scintillation at regional and global levels) and GNSS receiver and service products (nowcasting and forecasting of aviation related performance). The prototype IPS service provides 30 minute and 24 hour forecasts translated into GNSS user metrics such as position error. Users have direct access to data downloads and can easily access space weather conditions through the monitoring console. Notifications are sent when user-defined thresholds are crossed, and a forecast verification product is also provided. The unique quality of IPS is its capacity to provide information specifically targeted at GNSS users, and the next step in the project is to integrate IPS into the Galileo Service Center\footnote{https://www.gsc-europa.eu}. For the future of the service, it is necessary to increase awareness about the impacts of space weather on the routine operations of GNSS-based applications. 

The CME Scoreboard and CME arrival time and impact working team is part of the International Forum for Space Weather Capabilities Assessment\footnote{https://ccmc.gsfc.nasa.gov/assessment/}. The goals of the team include evaluating the status of CME arrival time and impact prediction, establishing community agreed metrics and events, and providing a benchmark from which future models can be assessed. Results were presented from the analysis of the NASA CCMC CME Scoreboard (Riley et al. 2018b), where 724 forecasts of CME shock arrival times were analysed between 2013 and 2018, during which 139 events took place. Forecasting errors in CME arrival time were calculated for each of the individual forecasting techniques submitted to the Scoreboard. Generally, precision (mean average error) in forecast time is +/- 13 hours, and the bias (mean error) is 2.8 hours on average. It was noted that forecasts were relatively consistent across the wide range of methodologies, and that SWPC, SIDC, and the ensemble CCMC model results provided the most accurate forecasts in this analysis. These particular models maintained the longest running forecasts in the CCMC database and were associated with long lead-times. Taking an average of all the methods also illustrated the value of super-ensembles as it performed best overall (Murray 2018).  It was recommended that official forecast centres work towards publishing forecasts in an internationally standardised, agreed-upon format (see for example Verbeke et al. 2018).     

The AUL framework tracks the progress of a project from the concept to prototype to operational stage and identifies best practices as to how research collaborators and users can work together. AULs help to communicate the status of projects and to advertise user needs and research capabilities. Three phases were defined, the first being Discovery and Viability, where fundamental research becomes applied. The second phase involves Development, Testing and Verification, where the focus is on finalising the development of a new state-of-the-art project, integrating the resulting tools into the identified applications, demonstrating the feasibility of the new product, and verifying the new system. Finally, Implementation and Integration into Operational Status, where the project is handed over and fully integrated into the new end-user’s application; this includes new verification efforts to determine how well the new application performs in the real-world, and the verification and use in the operational environment drives discovery of new science questions, problems, and new applications. Example projects were presented to illustrate the use of AULs. The website framework development is in progress which will indicate the status of different projects within the AUL framework and will link to CCMC’s verification activities.  

SWPC described operational experiences during the September 2017 events. In total, 123 alerts, watches, warnings and summary products were issued between 12 UTC 4th to 12 UTC 11th September 2017. This was five times more than issued during the entire previous month. Over the course of the whole month of September, 239 total hazard products were disseminated, 464 telephone calls were made to customers, and 18 media contacts were logged. SWPC said that it was difficult to keep on top of the warning emails sent to more than 250 thousand subscribers. SWPC also gave a useful summary of the sequence of severe events as described above, setting the scene for the panel discussion.    

\subsection{Panel discussion}
During the panel discussion, panellists from research and operations (J. Andries, K. Arsov, A. Halford, M. Hesse, R. Steenburgh and M. Temmer) considered their experiences during the September 2017 events. Some specific points were noted, for example power grid failures and GNSS degradation were reported in Australia (K. Arsov), while no feedback was reported from ionospheric users (R. Steenburgh). However, the panel discussion mainly focused on cross domain impacts of the September events. First, the impact of space weather on Caribbean hurricane relief efforts (Redmon et al. 2018) was discussed, when a geomagnetic storm caused disruption to HF radio communications used in the management of air traffic and emergency and disaster relief. The 6th September X9.3 and 10th September X8.2 flares led to rapid ionisation of the sunlit equatorial upper atmosphere, disrupting high frequency communications in the Caribbean region while emergency managers were attempting to provide critical recovery services during three devastating hurricanes. Both Hurricane Watch Net (HWN\footnote{https://www.hwn.org}) and the French Civil Aviation Authority (DGAC) reported issues during this period. For example, the X9.3 flare caused a near total communications blackout for most of the morning and early afternoon on the 6th September for ground operators. In addition, DGAC reported that HF radio contact was lost with one aircraft off the coasts of Brazil and French Guyana for about 90 minutes. The X8.2 flare disrupted HF communications on the ground on the 10th in the late afternoon for about 3 hours. 

The second cross domain impact discussed was experiences  of the Mexican Space Weather Service (SCiESMEX)\footnote{http://www.sciesmex.unam.mx/} (Gonzales-Esparza et al. 2018) during the combination of three different natural hazards: space weather, the category 2 hurricane Katia in the Gulf of Mexico (6th September) and two major earthquakes (Tehuantepec earthquake on the 7th and Puebla-Morelos on the 19th). This discussion received considerable interest from the audience as it was not as well-known as the Caribbean events. A geomagnetic storm occurred while civil protection authorities were assisting the affected population after the hurricane and Tehuantepec earthquake. Several posts on social media with fake information about solar storms as drivers of earthquakes created anxiety and fear in the population and the civil protection authorities had to react by sending information through social networks denying these claims. The combination of the three natural hazards highlighted the importance of reaction protocols against natural hazards (including the participation of SCiESMEX in providing reliable information on current solar activity), communication strategies and the use of social networks. 

The cross domain impact discussion highlighted the importance once again of education within the scientific, operational forecasting and end-user communities. In particular it was made clear that better communication is needed between the communities, with often a ‘valley of death’ disconnect between the basic science and applied output (A. Halford). Space Weather is still a new discipline under development which means that more research has to be done, but this fact should not stop or be used as an excuse for the dissemination and teaching of the current state-of-the-art (M. Temmer and J. Andries). 

The final question posed to the panellists was `Are we prepared for a severe space weather event?', \textit{we} being scientists and service providers. The consensus was that we are nearly there, and more so than the end-users since awareness levels tend to vary widely between sectors. It was noted however that such events can be complex and that more investment in research was required to further understanding. Panellists also noted that space weather forecasts and warnings depend heavily on available observations from scientific missions. To provide users with longer lead-time notifications would require future missions to replace end-of-life spacecraft and would require more operational missions to provide the required data.

The panellists agreed that space weather service providers are generally confident that they can provide adequate information during severe and extreme events. It was noted that during extreme events, faster CMEs can occur resulting in shorter lead-times of only about 13 hours between CME eruption and arrival at Earth. Also noted was that funding and ongoing research were certainly required to maintain current capabilities, particularly in the observation network, and to improve forecasts for the future.

\section{Commentary}
\subsection{Forecast accuracy}
Forecast accuracy is certainly an ongoing area of debate within the space weather community, and unsurprisingly emerged as a popular topic during the plenary sessions. Users expressed that longer lead-times are required than services are currently providing. For example, satellite operators discussed possible mitigation measures which could be taken with several weeks of notice of extreme events, while most operational centre forecasts tend to provide activity forecasts up to only four days ahead. Further to that, the accuracy of forecasts beyond even 24 hours is generally still in need of improvement (Sharpe \& Murray 2017).  This is an issue that is well known within the space weather community, and efforts are ongoing internationally to develop underpinning data provision and modelling in order to improve forecast capabilities. The increased use of cutting-edge techniques such as artificial intelligence as used by data scientists are also aiding these efforts, as noted in the services session. However, it was clear from panel discussions that this issue goes beyond scientific techniques, as end-users are asking to be provided services similar to those currently provided by weather forecasters (that they are comfortable with). The services session panel commented that the field of space weather forecasting is likely 30 years behind operational terrestrial weather forecasting, and the field may be able to learn from this community's successes and failures. Implementing operational modelling techniques frequently used in terrestrial meteorology, such as ensembles and data assimilation (Murray 2018; Henley \& Pope 2017), has for example proved helpful whilst the community waits for research improvements, as has designing familiar visualisation outputs for users. Discussion within the sessions made clear that user requirements were not yet fully met by the current generation of services in terms of accuracy and lead-time.

\subsection{Preparedness}
There was discussion on how prepared user communities were for a severe event. At national level, some countries have developed strategies for managing the risk of space weather, for instance the UK strategy as mentioned above and the US National Space Weather Strategy and Action Plan (White House 2019). Such plans are intended to improve the nation’s preparedness for space weather events. Understanding in preparedness is improved by further understanding the impacts of space weather; studies such as that in the US on space weather benchmarks (White House 2018) provide estimates of the characteristics of a 1-in-a-100 year event. The US benchmark study provides estimates for the theoretical maxima for such an event for five phenomena: induced geo-electric fields, ionising radiation, ionospheric disturbances, solar radio bursts and upper atmosphere expansion. Such studies provide the research community with an understanding of what is required, for example in modelling extreme events, thus providing a sound basis for improved predictions and services.

Gaps in service capabilities and solutions to these gaps were regularly discussed throughout the sessions. A global roadmap for 2015-2025 (Schrijver et al. 2015) commissioned by the Committee on Space Research (COSPAR) and International Living With a Star (ILWS) provides an assessment of the field of space weather research taking into account end-user viewpoints, and identifies gaps in scientific understanding, modelling capabilities and observational coverage. The roadmap makes recommendations which will improve understanding and forecasts of space weather. For example, one of the highest priority research areas identified to improve space weather information is to quantify active region magnetic structure to model nascent CMEs. The World Meteorological Organisation (WMO) fosters the development of best practices for space weather products and services (WMO 2016), for example through building on the experience of International Space Environment Service (ISES)\footnote{http://www.spaceweather.org/} Regional Warning Centres to improve accuracy, reliability, interoperability and overall cost-efficiency of provision of services. Within Europe the ESA SSA Programme is developing a European space weather service system, building upon existing expertise and federating more than 40 institutes and organisations from across Europe in order to provide timely and reliable space weather information to service users. Good practices have been adapted from weather services, and were discussed in the AUL presentation, such as the engagement of users throughout the process of defining to operationally implementing a space weather tool.  

\subsection{Engagement and Education}
Regarding user engagement a topic of contention within the sessions was the definitions of ‘extreme/severe/strong’ events, as their interpretation notably varied within different end-user groups. In particular there was some confusion in the discussion as to why the September 2017 events were classed as ‘severe’. As mentioned above, there was a ‘severe’ geomagnetic storm but only ‘strong’ radio blackout and solar particle events at most. And so, when referring to the type of an event, the parameter or parameters should always be referred to which make it such an event. There is clearly a need for increased communication between service providers and user groups to avoid confusion surrounding space weather forecast definitions, and also increased education to understand the meaning of such terms.

Both panel discussions noted the importance of education, throughout the community from researchers to users, for raising awareness of current activities in the space weather community. The need for introductory space weather courses as a first step was highlighted, since it requires substantial resources to tailor sector-specific and impact-specific content considering the vast array of effects across the Sun-Earth connection. Introductory courses such as those provided at the start of ESWW\footnote{see, e.g., http://www.stce.be/esww13/tutorial.php} have proved helpful in bringing a variety of users together to educate them on impacts as well as forge new connections. Relating to industry education, three space weather schools for engineers have been organised in the framework of ESWW. The fourth was unfortunately cancelled\footnote{ http://www.stce.be/sw4e/} due to the lack of participants. The difficulty in holding such tailored courses is that some companies may not be sufficiently aware of space weather impacts to their industry and so do not prioritise funding and time to train an engineer for several days on space weather hazards. To help increase attendance on these courses the potential benefits to the industries concerned could be impressed on budget holders in plain terms and in terms of economic value with examples given of industries which have already benefited; industries could also be involved in planning the program ensuring relevance and encouraging engagement; the program content has to be informative without focusing on a very specific set of users. 

Some efforts presented in the services session, such as AULs, also showed how essential it is to educate researchers into the difference between research software and operational capability. For example, operational services may only support certain programming languages such as Python and Fortran, and model output must be produced in a timely manner to provide a forecaster with useful information, with backup systems available in the case of faults. Clearly defining best practice in the development process, from proper testing, documentation, and performance assessments, will help researchers undertake the steps needed to transition a basic research idea into application. It was also noted in the panel discussions that there is an increasing awareness amongst researchers that their main users are the forecasters, the brokers with end-users in industry. This has helped focus science efforts into tackling long-standing research challenges, whilst operational centres can focus on tailoring that science output into sector specific guidance.

The need for improved education led discussion back to the need for increased engagement, particularly with underrepresented end-user groups. The example of the rail industry arose during the users session, which has not yet established space weather guidelines to the extent of, for example, the aviation or power industries. A workshop on space weather and rail was held in the UK in September 2015 (Krausmann et al. 2015) to raise awareness of this topic and to further explore the possible vulnerability of rail systems, for example the potential for signalling anomalies to occur during a space weather event. It was noted during the users session that the rail sector needs to identify its vulnerabilities and subsequently the early-warning requirements necessary to better protect its assets. Tracking the efforts underway in this new sector could prove to be helpful for other areas that have not yet undertaken space weather impact studies.

\section{Conclusion}
Space weather service providers highlighted that it was essential to engage with users to understand requirements thus enabling the development of products and services which are fit for purpose. It was noted that service providers benefit greatly by working with smaller groups of particularly engaged users. Overall, the two sessions highlighted several cases where space weather users and service providers were generally working well together and that there was plenty of opportunity to improve. For example, the space sector will typically make use of information about the space environment and its variability both during the design and operational phases of a mission but transport sectors such as rail could benefit if services could give them a greater understanding of potential impacts.

The main improvements needed to services, arising from the discussions, were longer forecast lead-times and accuracy, training in space weather, and greater engagement between service providers and users. In addition, service providers commented on challenges frequently encountered in maintaining funding streams to support continued service provision beyond the conclusion of project-based development funding and in securing project-based funding for research to operations. Availability of space weather observations was also discussed, both in terms of the importance of maintaining currently available observations and in providing new complementary observations geared towards improving space weather service capability, such as the Lagrange mission to L5 currently under study as part of the ESA SSA Programme.

\section{Acknowledgements}
The authors would like to thank the contributors to the ESWW15 plenary sessions: J. Andries, K. Arsov, E. Guyader, A. Halford, M. Hesse, R. Leussu, L. Lochman, S. Magdaleno, M. Mays, J. Mun, D. Pitchford, R. Steenburgh, M. Temmer, D. Zamora and BEIS, along with the ESWW program committee. The authors would also like to thank the reviewers of the manuscript for valuable feedback. S.A.M. is supported by the Irish Research Council Postdoctoral Fellowship Programme and Air Force Office of Scientific Research award number FA9550-17-1-039.

\section{Appendix}
A number of talking points were provided to presenters and panellists to encourage discussion during the panel sessions:

\vspace{5mm}
\noindent For the users session:
\begin{itemize}
\item This is a cross domain session themed around how events observed and responses in 2017 might scale in the case of an extreme Carrington-like event
\item What information is expected from a space weather service provider during a severe event and an extreme event?
\item What are the current practices and experiences in utilising space weather products and services?
\item What are the high priority needs for actionable space weather information? 
\item What could be improved in the available space weather information?
\item What direct space weather impacts on infrastructure have you suffered?
\item How prepared are you for a significant space weather event?
\item Where would you go for information during a space weather event? e.g. space weather information sources, government, media outlets, civil contingencies, domain authorities (such as Air Traffic Control)
\item Do you have an action plan? Have you practiced an action plan?
\item Are other user domains you rely on prepared?
\item How prepared do you think society is for a significant space weather event? What are the gaps in society’s overall response capability?
\end{itemize}

\vspace{5mm}
\noindent For the services session:
\begin{itemize}
\item As a service provider, what experiences did you have during Sept 2017 events?
\item Was September 2017 worthy of the name ‘severe events’?
\item Can the user needs (alerts and forecasts) be met? For September 2017 and Carrington events?  For all-clear periods?
\item Provide examples of services which give clear and simple predictions?  Can we do more? 
\item How can effective user education be provided?
\item Can service providers more effectively engage the transport sector?  e.g. provide evidence of solar activity impact on rail
\item Are we prepared for cross domain impacts? e.g. hurricane/pandemic flu occurring at the same time as space weather 
\item What developments are required in current forecasts to ensure full capability of services during severe and extreme events? Where are the gaps in capability?  
\item What are the solutions to these gaps? e.g. an L5 mission, improvement in access to operational and real-time data, more research, greater understanding between service providers and users
\item Service providers need to understand user limitations (e.g. satellite technology risk thresholds and anomalies) but it can be difficult for the user to share information – how can we overcome this challenge?
\item How will service providers meet future user needs? e.g. supporting autonomous vehicles and the launch of thousands of new satellites 
\end{itemize}

\end{document}